\documentclass[summary]{gass2017}

\usepackage{amsmath}
\usepackage{lipsum}
\usepackage{graphicx,epstopdf}
\usepackage{amsfonts}
\usepackage{amssymb}
\usepackage{dcolumn}
\usepackage{amsmath}
\usepackage{soul}
\usepackage{color}

\usepackage{amssymb}
\usepackage{amsfonts}
\usepackage{amsmath}

\def\e{\begin{equation}}
\def\f{\end{equation}}
\def\=#1{\overline{\overline #1}}
\def\*#1{\overline{\overline{\overline #1}}}
\def\-#1{{\bf #1}}
\def\_#1{{\bf #1}}
\def\l#1{\label{eq:#1}}

\newcommand{\halpha}{\hat{\alpha}}

\newcommand{\br}{\mathbf{r}}

\newcommand{\bk}{\mathbf{k}}

\title{Functional metasurfaces: Do we need normal polarizations?}

\author{
{Sergei A. Tretyakov*\affref{ref1}, 
Do-Hoon Kwon\affref{ref1}\affref{ref2}}, Mohammad Albooyeh\affref{ref3}, 
and Filippo Capolino\affref{ref3}}

\affiliation{%
  \aff{ref1}{Aalto University, Finland}
  \aff{ref2}{University of Massachusetts Amherst, Amherst, Massachusetts, USA}
  \aff{ref3}{University of California, Irvine, California, USA}
  }

\begin{document}

\maketitle

\begin{abstract}

We consider reciprocal metasurfaces with engineered reflection and transmission coefficients and study the role of normal (with respect to the metasurface plane) electric and magnetic polarizations on the possibilities to shape the reflection and transmission responses.  We demonstrate in general and  on a representative example that the presence of normal components  of the polarization  vectors does not add extra degrees of freedom in engineering the reflection and transmission characteristics of metasurfaces. Furthermore, we discuss advantages and disadvantages of equivalent volumetric and fully planar realizations of the same properties of functional metasurfaces.

\end{abstract}

\section{Introduction}

Composite layers with electrically negligible thickness and engineered electromagnetic properties, called \emph{metasurfaces,} are actively developing as effective means to control reflection and transmission of electromagnetic waves, see e.g the review paper \cite{review}.
In most works, the focus is on engineering
{induced currents \emph{tangential to the surface}.}
This approach is based on the Huygens principle, which tells that equivalent tangential currents on a closed surface fully determine the fields inside the volume enclosed by this surface.
Although metasurfaces have a finite thickness and 
{ polarizations normal to the surface are also induced}, 
it appears that  their role on the metasurface performance is not yet properly understood. In an interesting recent paper \cite{Grbic} it is stated that ``adding longitudinal (normal) surface currents significantly expands the scope of electromagnetic phenomena that can be engineered with reciprocal materials''. This is a very counter-intuitive conclusion, since it is generally believed that setting only tangential currents is enough to fully control the reflected and transmitted fields, and here  we consider this issue in due detail. 

In this presentation we discuss the role of normal polarizations on the  far-field response of general reciprocal metasurfaces. We consider example metasurfaces which have interesting asymmetric properties  due to the presence of normal polarizations (like the structures discussed in \cite{Grbic}) and show that it is in fact possible to realize the same response using equivalent metasurfaces which support only tangential polarizations. Finally, we discuss practical advantages and disadvantages of  metasurfaces of both types.

\section{Equivalency of effects due to normal and tangential polarizations}

\begin{figure}[h!]
\centering
 \includegraphics{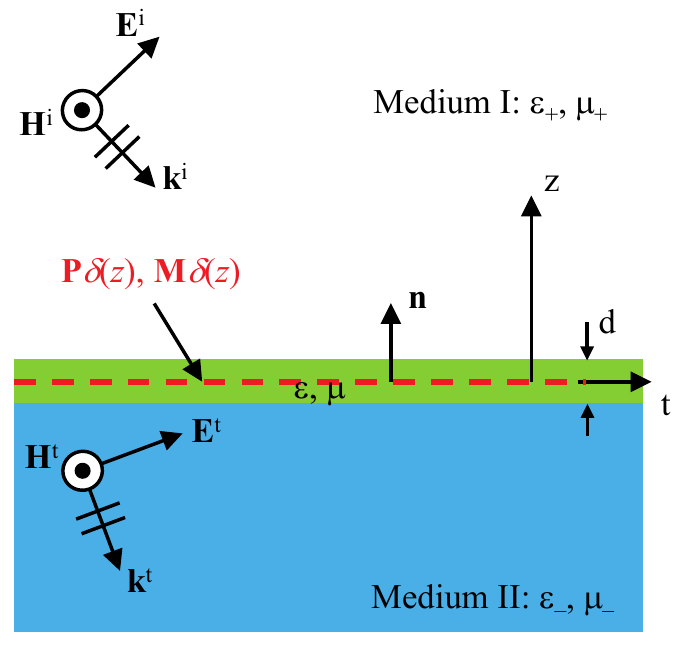}
\caption{Geometry of a metasurface composed of an electrically dense planar array of arbitrarily polarizable inclusions 
({electric and magnetic} polarization vectors averaged 
over the unit cell are denoted as 
$\mathbf{P}$ and $\mathbf{M}$, {respectively}). 
The unit vector \textbf{n} is normal to the metasurface.
{$\epsilon_{\pm}$ and $\mu_{\pm}$ are the relative permittivity and permeability of the corresponding medium.}
}
\label{FIG_SCHtD}
\end{figure}

Let us consider a general reciprocal metasurface, illustrated in Fig.~\ref{FIG_SCHtD}. Surface-averaged {electric and magnetic}
polarization vectors $\-P$ and $\-M$
can have arbitrary directions with both tangential and normal components. In the limit of negligible thickness ($d\rightarrow 0$) the $z$-dependence of the induced polarizations can be replaced by a Dirac delta-function.

Due to  the presence of surface polarizations, tangential fields are not continuous across the metasurface and exhibit jumps  \cite{Idemen,Kuester,Annalen}:
\begin{eqnarray}\l{STBC}
  \nonumber \-E^+_\textrm{t} - \-E^-_\textrm{t} &=& j \omega {\-n} \times {{\-{ M}}_\textrm{t}} - \-\nabla_\textrm{t} {{{ P}}_\textrm{n} \over {\epsilon}}, \\
  {\-n} \times \-H^+_\textrm{t} - {\-n} \times \-H^-_\textrm{t} &=& j \omega {{\-{ P}}_\textrm{t}} + \-\nabla_\textrm{t} \times {\-n} {{{ M}}_\textrm{n} \over {\mu}}.
\end{eqnarray}

We can immediately see that the tangential gradient of the normal component of electric polarization produces the same effect as tangential magnetic polarization, and, similarly, spatially varying normal component of magnetization acts equivalently to tangential electric polarization.  Therefore, we can introduce  new vectors
\begin{equation}\l{defMs}
 {\_M}_\textrm{te} = {{\_{ M}}_\textrm{t}} +\_n\times  \nabla_\textrm{t} {{{ P}}_\textrm{n} \over {j\omega \epsilon}},
\end{equation}
and
\begin{equation}\l{defPs}
{\_P}_\textrm{te}= {{\_{ P}}_\textrm{t}}  + \-\nabla_\textrm{t} \times {\-n} {{{ M}}_\textrm{n} \over {j\omega \mu}},
\end{equation}
which, respectively, represent the \textit{total}
{equivalent} tangential magnetic and electric surface polarization densities contributing to the discontinuity of the tangential electric and magnetic fields. These simple considerations suggest that any effect of normal polarizations on the surface-averaged tangential electric { and magnetic}
fields can be reproduced by properly engineered tangential polarizations.
Since the {tangential fields averaged over the unit cell}
on the surface fully determine far-field reflected and transmitted fields, this equivalency applies to arbitrary engineered reflection and transmission properties.

\section{Example of equivalent tangential polarizations}

\begin{figure}[h!]
\centering
 \includegraphics{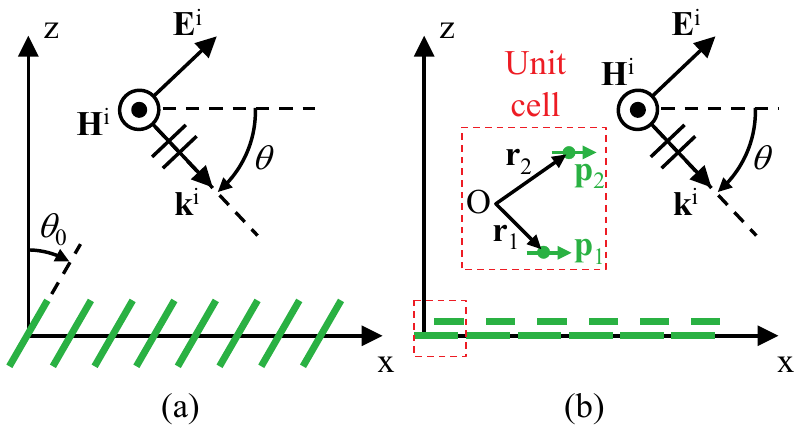}
\caption{Schematic of two physically different metasurfaces with equal responses to the same excitation.  In (a), there are induced electric moments with tangential and normal components while in (b) there are both electric and magnetic moments which are only tangential to the metasurface plane.}
\label{FIG_SCH}
\end{figure}

As a representative and illustrative example we consider a metasurface shown in Fig.~\ref{FIG_SCH}(a). The unit cell of the metasurface is formed by an electric dipole (shown by a green line) which is tilted with respect to the metasurface plane. The unit cell size is subwavelength, but we assume that the dipole can have resonant response, for example due to a bulk load in its center.
This metasurface has an interesting property: 
for illuminations at the angles 
{$\theta=\pi/2+\theta_0$} and 
{$\theta=\pi/2-\theta_0$} 
the reflection coefficient vanishes.
Indeed, 
{if $\theta=\pi/2+\theta_0$},
there is no reflection simply because the electric field is orthogonal to the dipoles and there are no induced currents.
If {$\theta=\pi/2-\theta_0$}, 
there is also no reflection because the system is reciprocal. Although the reflection coefficient is the same for these two illuminations, the transmission coefficient is not symmetric.
For illumination at {$\theta=\pi/2+\theta_0$} 
the transmission coefficient is unity, since there are no induced currents, 
but for illumination at {$\theta=\pi/2-\theta_0$} 
it can be tuned to any desired (physically allowed) value varying the dipole load impedance (including full absorption or full control over the transmission phase). From the reciprocity theorem it follows that the transmission coefficients at 
$\theta=\pi/2-\theta_0$ and $\theta=3\pi/2+\theta_0$  or $\theta=\pi/2+\theta_0$ and $\theta=3\pi/2-\theta_0$
(illuminations from the opposite sides of the array) have the same asymmetry property.

This asymmetry of transmission at 
{$\theta=\pi/2-\theta_0$ and $\theta=3\pi/2+\theta_0$} 
was studied in \cite{Grbic} as a property enabled by non-zero normal polarizations induced in the metasurface inclusions. Let us show that the same properties can be realized using a metasurface which is polarizable only in the tangential plane. An appropriate geometry of inclusions is shown in Fig.~\ref{FIG_SCHtD}(b). 

The unit cell consists of two different {tangential}
electric dipoles, shifted with respect to each other as shown in the picture. Obviously, only tangential polarizations exist in this case. 
In terms of the effective polarizabilities of the dipoles in periodical arrays \cite{modeboo}, the induced dipole moments can be found as functions of the incident electric field:
\begin{align}
p_1 &= \halpha_1[E^\textrm{i}_\textrm{x}(\br_1)+\beta p_2],\label{p1}\\
p_2 &= \halpha_2[\beta p_1+E^\textrm{i}_\textrm{x}(\br_2)].\label{p2}
\end{align}
Here, $\halpha_1$ and $\halpha_2$ are the effective polarizabilities of each dipole in the corresponding array which are defined as \cite{modeboo}
\begin{equation}
\halpha_1 = {1\over {{1\over\alpha_1}-\beta_1}},\qquad 
\halpha_2 = {1\over {{1\over\alpha_2}-\beta_2}},\label{al2}
\end{equation}
and the interaction constants of the two dipole arrays are denoted as $\beta_{1,2}$.
The interaction constant $\beta$ defines the field created by one of the two dipoles arrays at the positions of the dipoles in the other array. $\alpha_{1,2}$ are the polarizabilities of single dipoles in free space.

Solving the above equations, the transmission coefficient can be found:
\begin{equation}
T=1 {-} j\omega\eta|\sin\theta|
\frac{\halpha_1+\halpha_2+2\halpha_1\halpha_2\beta
\cos[\bk\cdot(\br_2-\br_1)]}{2S(1-\halpha_1\halpha_2\beta^2)},
\label{t:strips}
\end{equation}
where $\_k={\_k^{\rm i}}$ 
is the wave vector of the incident plane wave,  $\eta$ is the wave impedance of the surrounding space, and $S$ is the unit cell area.

Inspection of (\ref{t:strips}) reveals that it is possible to realize the same property of asymmetric transmission (and zero reflection) as observed in arrays of tilted dipoles if the following conditions are satisfied.   First, a non-zero
interaction constant $\beta$ is required. Next, it is required that
$\cos[\bk\cdot(\br_2-\br_1)]$ must be different for illuminations at 
{$\theta=\pi/2-\theta_0$ and $\theta=3\pi/2+\theta_0$}, 
which is possible only 
for asymmetric strip arrangements in the $x$-direction
(i.e., $\br_2-\br_1$ not parallel to the $z$-axis).
Obviously, single planar structures do not
permit asymmetric transmission. 

In the presentation we will show numerical results illustrating the same properties of this example array of only tangentially polarizable particles and the array of tilted dipoles where also normal polarizations are present.

\section{Conclusion}

We have shown that  an arbitrary  wave manipulation by the metasurface which is possibly enabled by engineering both tangential and normal induced polarization currents can be obtained also by using only equivalent tangential induced currents; hence, removing the need for induced normal polarizations in a metasurface. 
In practical terms, this conclusion is important because it shows that most general reciprocal functional metasurfaces can be realized using only planar, easily manufacturable structures. 

So, is the answer to the question ``Do we need normal polarizations?'' negative? We would not make such a conclusion. Indeed, even in the presented examples, the version of tilted dipoles has important advantages. The property of zero reflection at the angle 
{$\theta=\pi/2-\theta_0$} is extremely robust 
with respect to the frequency and dipole loads, since it is ensured by the fixed pattern null along the dipole {axis}.
In the planar version, the reflection zero requires a careful balance of amplitudes and phases of the currents induced in the two dipoles, which can be achieved only at one particular design frequency. Likewise, engineering of transmission phase in the array of tilted dipoles is achieved simply by setting the proper values of the dipole load reactances, while in  the planar version numerical optimization is needed to synthesize the desired phase response. Thus, in practice there is a compromise between easy manufacturing and robust response.



\end{document}